\documentstyle[12pt]{article}
\begin{document}

\def\bce{\begin{center}}
\def\ece{\end{center}}
\def\beq{\begin{eqnarray}}
\def\eeq{\end{eqnarray}}
\def\ben{\begin{enumerate}}
\def\een{\end{enumerate}}
\def\ul{\underline}
\def\ni{\noindent}
\def\nn{\nonumber}
\def\bs{\bigskip}
\def\ms{\medskip}
\def\tr{\mbox{tr}}
\def\wt{\widetilde}
\def\wh{\widehat}
\def\brr{\begin{array}}
\def\err{\end{array}}
\def\dsp{\displaystyle}
\def\eg{{\it e.g.}}
\def\ie{{\it i.e.}}

\vspace*{-10mm}
 %\hfill IEEC/CSM-01-81

 \hfill math-ph/0109006

 \hfill August 30, 2001
\thispagestyle{empty}

\vspace*{4mm}
 
\begin{center}

{\LARGE \bf On recent strategies proposed for proving 
the Riemann hypothesis}

\vspace{4mm} 
\medskip

{\sc E. Elizalde\footnote{E-mail:
 elizalde@math.mit.edu \ elizalde@ieec.fcr.es \
\ http://www.ieec.fcr.es/recerca/cme/eli.html \\  On leave from:
Instituto de Ciencias del Espacio (CSIC)  \&
Institut d'Estudis Espacials de Catalunya (IEEC/CSIC),
Edifici Nexus, Gran Capit\`a 2-4, 08034 Barcelona, Spain}} \\
Department of Mathematics,  Massachusetts Institute of Technology\\ 
77 Massachusetts Avenue,
Cambridge, MA 02139-4307

{\sc  V. Moretti\footnote{E-mail: moretti@science.unitn.it}} \\
Dipartimento di Matematica,
Universit\'a di Trento, Via Sommarive 14, I-38050 Povo (TN)\\

 {\sc  S. Zerbini\footnote{E-mail: zerbini@science.unitn.it}} \\
Dipartimento di Fisica, Universit\'a di Trento \\ \&
INFN Gruppo Collegato di Trento,  Via Sommarive 14, I-38050 Povo (TN)\\

\vspace{6mm}
 
{\bf Abstract}
 
\end{center}
We comment on some apparently weak points in the novel
strategies recently developed by various authors
aiming at a proof of the Riemann hypothesis. After noting the
existence of relevant previous papers where similar  tools have
been used, we refine some of these strategies. It is not clear at the
moment if the problems we point out here can be resolved
rigorously, and thus a proof of the RH be obtained, along the lines proposed.
However, a specific suggestion of a procedure to overcome the 
encountered difficulties is made, what constitutes a step towards this goal.

\vfill

MSC-class: \ 11M26; 30B40; 14G10; 46E20.

%\vspace{2cm}

\newpage

As is well known, the Riemann zeta function, $\zeta_R$,  
has zeros at all negative even integers, these are called
{\em trivial} zeros. The {\em Riemann hypothesis} \cite{R} conjectures  
 that all of the remaining zeros, i.e.,  the {\em nontrivial}
ones, have real part equal to $\frac{1}{2}$.
There is no doubt that the proof of the Riemann hypothesis is one of the most
outstanding open problems in Mathematics. Suffice to say that it is the only
one problem that has been transferred from the famous list of Hilbert
of AD 1900 to the new list elaborated under the auspices of the Clay 
Institute in AD 2000.

Recently, a  beautifully simple approach towards 
the resolution ---in the positive sense--- of the Riemann hypothesis 
(also called sometimes the Riemann conjecture) has been elaborated in work by
M. Pitk\"anen \cite{pitka1}, and by C. Castro,  A. Granik, and J. Mahecha
\cite{cc1,cgm1}. It is an interesting approach, which involves 
 powerful techniques of zeta function regularisation \cite{zfr1},
together with the more commonly employed methods that make use of the
correlation function of the distribution of the non-trivial zeros of $\zeta_R$ 
and of the statistical fluctuations of a chaotic system related with
them \cite{chao1}. That these authors have been able to connect both types
of strategies, which apparently seem to be completely unrelated and far 
from each other, is already a remarkable achievement.

In essence, two are the main ideas involved in the strategies developed by
the above authors: first, to formulate the problem of finding the non-trivial 
zeros of $\zeta_R$ in terms of orthogonality properties of some functions
belonging to a Hilbert space and, second, the already mentioned one of
regularisation of the scalar product through analytical continuation 
by means of the zeta function. Although (surprisingly) not quoted by
these authors, the first (extremely nice) idea has been pursued in the
mathematical literature for a number of years (see \cite{sp1} for just a few 
quite recent references). Concerning the second issue, as specialists in the
field we should point out the following.  

The power and usefulness of the analytical continuation implied  by the 
zeta regularisation method is not always without danger \cite{zfr0}. Quite 
on the contrary, 
a big amount of erroneous calculations performed by using this method have 
been reported in the literature \cite{zfr0}, and the moral is that 
one must be extremely cautious at every step, when applying the procedure. 
In particular, a most common source of error has its roots in the loss of
linearity introduced by the zeta function. Specifically, the 
zeta trace of a differential (more generally of a pseudodifferential)
operator is {\it not} linear \cite{eejhep1}, e.g.,
\beq
\tr_\zeta (A+B) \neq  \tr_\zeta A + \tr_\zeta B,
\eeq
what leads immediately to the appearance of a generic multiplicative 
anomaly of the zeta determinant of the operator (this holds even for {\it 
commuting} operators, $[A,B] =0$) \cite{nca1} 
\beq
{\det}_\zeta (A \, B) \neq {\det}_\zeta A \ {\det}_\zeta B,
\eeq
which can be  nicely expressed in terms of the Wodzicki (or 
non-commutative) residue \cite{wod1}. This is, let us repeat, a 
clear consequence of the loss  of linearity that occurs when performing 
the analytic continuation \cite{eejhep1}.

We now go back to the first fundamental issue in all the (related) approaches 
introduced in 
\cite{pitka1,cc1,cgm1}, namely, the definition of a convenient Hilbert space 
on the half-line $(0,\infty)$ (see also \cite{sp1}), with the scalar product
\beq
\langle\psi\vert\phi\rangle  =    \int_0^\infty \frac{dt}{t}  \,
\psi(t)^*\phi(t) ,
\eeq
and consider therein the eigenfunctions, $\psi_s(t)$
with complex eigenvalues  $s\in C$,
of a
convenient differential operator, $D$ with respect to the variable $t$,
\beq
D\, \psi_s = s\, \psi_s, \qquad s\in C.
\eeq
The operator $D$ acts on complex-valued functions defined on
$(0,+\infty)$.

The scalar product
\beq
\langle\psi_{s_1}\vert\psi_{s_2}\rangle = \int_0^\infty \frac{dt}{t} \,
\psi_{s_1}(t)^*\psi_{s_2}(t) 
\eeq 
yields a  finite result  as far
as the eigenvalues $s$ are constrained to a domain that corresponds,
in all these approaches, to the region delimited by the abscissa of convergence
of the zeta function of the relevant operator \cite{zfr1}. The existence of 
the nontrivial zeros of the Riemann zeta function {\it only} on the 
critical line 
Re $z =1/2$ is then reduced (iff) to an orthogonality condition of the 
eigenfunctions with respect to the above scalar product, which definition must be 
necessarily extended {\it beyond} the domain of absolute convergence in the 
variables $s_1,s_2$.

This is most naturally done by using zeta function regularisation in
its most basic version, that is, by extending on the rhs a 
convenient  representation
of the zeta function ---which coincides with the integral at the rhs of the 
definition of the scalar product at the aforementioned domain of absolute 
convergence--- to the
rest of the complex plain, that is \cite{cgm1}
\beq
\langle\psi_{s_1}\vert\psi_{s_2}\rangle \sim \zeta_R (2(2k-s_{12})/l).
\label{5a}
\eeq 
Here $k$ and $l$ are real numbers, that can be chosen conveniently, and 
$s_{12}=s_1^*+s_2=x_1+x_2+i(y_2-y_1)$. The 
problem is now that the ``analytically continued scalar product'' 
thus obtained ceases in fact to be a scalar product. In particular, it 
is no more {\it bilinear}, aside from the additional problem of not
being positive definite. This last condition, positivity, can be 
apparently restored (by switching in a convenient global constant), at least 
in the domain of the complex plane relevant for the final 
argument, the one that leads to the proof of the Riemann hypothesis. However,
the loss of linearity ---similar to the one that gives rise to the 
multiplicative anomaly of the determinant, as explained before--- 
is {\it not} so easy to fix and will require some deeper
 investigation (in particular, the product 
$\langle\psi_{s}\vert\psi_{1-s}\rangle$ is needed to be given sense as 
a {\it scalar} product). Unless we manage to have in fact
 a scalar product in the
analytically continued domain, all further references to the concept 
and properties of {\it orthogonality}, in its one-to-one relation with 
the presence of a zero (at the critical line Re $s=1/2$) of the 
Riemann zeta function ---that appears on the rhs of Eq. (\ref{5a})--- 
simply stop to make sense. This precludes the obtaintion of a 
mathematical proof of the elusive Riemann conjecture.

A first line of thought in order to resolve the issue dealt with here
 might go through a deeper understanding of the space of eigenfunctions 
and, eventually, also through the construction of a different operator $D$,
in an attempt to restore the linearity of the scalar product in the 
analytically continued domain relevant for the proof.
To this end,  we would like to add here some specific considerations,
 along the lines of Refs. \cite{pitka1,cgm1,sp1}, {\it but} trying 
always to deal with a proper scalar product in a suitable Hilbert space.

Let us consider the  Hilbert space  $L_2(0,\infty)$ referred to the usual
scalar product
\beq
(f,g)=\int_0^\infty dT f^*(T)g(T)\,.
\eeq
induced by Lebesgue's measure. Consider then the differential operator
(generalized annihilation operator)
\beq
A=\frac{d\:\:}{dT}-\frac{1}{2\omega(T)}\frac{d \omega}{d T}\,,
\eeq
defined on the domain ${\cal D}(A)$ of  elements $\phi \in C^1(0,\infty)
\cap L_2(0,\infty)$  with  $A\phi \in L_2(0,\infty)$.
${\cal D}(A)$ is a  dense linear manifold in $L_2(0,\infty)$.
The $\omega$ above is  a smooth non-negative function which we shall 
specify shortly.
Unfortunately, despite ${\cal D}(A)$ being dense, it turns out that
 $iA$ (as well as 
$-i\frac{d\:\:}{dT}$, because of the lack of a boundary condition)
fails in fact to be  symmetric and thus no self-adjoint extensions 
are possible. This entails that, at this step, the
  usual tools of  spectral analysis  for self-adjoint operators  cannot 
be directly implemented.

Notwithstandingthat, it is easy to show that
\beq
\phi_s(T)=e^{sT/4}\sqrt{\omega(T)}\,,
\eeq
is a formal eigenfunction for  $A$ corresponding to
the eigenvalue $s/4$,  $s$ being an  arbitrary complex number.
Moreover, for our aims, the function $\omega(T)$ is conveniently chosen as
\beq
\omega(T)=\sum_{n=1}^\infty e^{-\pi n^2 \exp T}\,.
\eeq
As a result, one gets  that, for each $s\in C$, $\phi_s \in {\cal D}(A)$,
  since  $||\phi_s|| <\infty$ and no boundary condition has been 
imposed at $T=0$. This means   that {\em all} of the
$\phi_s$ are  {\em proper} eigenvector of $A$.
Notice that the imposition of  boundary conditions in the definitions of 
${\cal D}(A)$ would select only {\em some} of the above formal 
eigenfunctions as  proper eigenfunctions.
This fact could  be useful in further investigations. However, we want 
here to focus attention on the simplest case only.

By performing the change of variables $t=\exp T$,   one finds that   
\beq
(\phi_s,\phi_z)=F(\frac{s^*+z}{2})\,,
\eeq
for all  pairs $s,z\in C$.
Above, we have introduced the function $F(s)$, which is
 analytic in the whole complex plane and defined by
\beq
F(s)=\int_1^\infty dt\ t^{s/2-1}\omega(t)\,.
\eeq
We also have the relation
\beq
(\phi_s,\phi_z) = (\phi_{s-w^*},\phi_{z+w})\,,        
\label{prod}
\eeq
for all $s,z,w \in C$. Thus,
\beq
(\phi_s,\phi_z) = (\phi_0, \phi_{z+s^*})\,,          
\label{prod1}
\eeq
where $\phi_0$ is the zero mode of $A$. 

On the other hand, in the theory of the Riemann zeta function,  the following
relation is well known to be valid in the whole complex plane
\beq
Z(s)=\pi^{-s}\Gamma(s/2)\zeta_R(s)=\frac{1}{s(s-1)}+F(s)+F(1-s)\,.      
            \label{Z}
\eeq
Since $\Gamma(s/2)$ has no zero, $s$, with $\mbox{Re } s =1/2 $, the  
relation above and (\ref{prod}) allow us
to re-state Riemann's hypothesis in terms of eigenfunctions of the 
operator $A$ and
a (well posed) Hilbert scalar product as follows. 

First, we have the

\noindent{\sf Proposition.}
{\em For the complex numbers $z\neq -4n$, $n=1,2,3\ldots$, consider
\beq
(\phi_0, \phi_{z} + \phi_{2-z}) = \frac{4}{z(2-z)}.                     
        \label{R}
\eeq
The complex number  $s= z/2$ is a zero of $\zeta_R$
if and only if  $z$ satisfies {\em (\ref{R})}.}

From Eq. (\ref{R}) and using Eq. (\ref{prod})  we also get the

\noindent{\sf Corollary.}
{\em The value
  $s=z/2=1/2+iy$  
(where $y$ is real) is a non-trivial zero of $\zeta_R$
iff
\beq
(\phi_0, \phi_{1+i2y} + \phi_{1-i2y}) = \frac{4}{1+ 4y^2}             
         \label{R'}\,.
\eeq
}
This is, in our view, a valid step towards the complete resolution of 
the problem, although the identification of the 
operator as corresponding to a precise (physical) process is still lacking.

\bs

\noindent{\bf Acknowledgments}

EE is indebted with the Physics Department, Trento University, and the
Mathematics Department, MIT, for the very warm hospitality.
This investigation has been supported by DGI/SGPI (Spain), project
BFM2000-0810, by CIRIT (Generalitat de Catalunya),
contract 1999SGR-00257, and by the program INFN (Italy)--DGICYT (Spain).

\vspace{5mm}

%\newpage


\begin{thebibliography}{99}

\bibitem{R} B. Riemann, {\it \"Uber die Anzahl der Primzahlen unter einer 
gegebenen Gr\"osse},
Monat.    der K\"onigl. Preuss. Akad. der Wissen. zu Berlin aus der Jahre 
1859 (1860) 671-680; 
also  Gesammelte math. Werke  und wissensch. Nachlass. 2 {\em Aufl}.  
1892, 145-155. (See also the documented web page by M. Watkins, 
www.math.ex.ac.uk/$\sim$mwatkins/zeta/).

\bibitem{pitka1} M. Pitk\"anen, {\it Riemann hypothesis and superconformal
invariance}, math.GM/0102031.

\bibitem{cc1} C. Castro, {\it On p-adic stochastic
dynamics, supersymmetry and the Riemann conjecture}, physics/0101104.

\bibitem{cgm1} C. Castro, A. Granik, and J. Mahecha, {\it 
On SUSY-QM, fractal strings and steps towards a proof of the Riemann
hypothesis}, hep-th/0107266.

\bibitem{zfr1}
 E. Elizalde, S.D. Odintsov, A. Romeo, A.A. Bytsenko, and S. Zerbini,
{\it Zeta Regularization Techniques with Applications}
(World Scientific, Singapore, 1994);  E. Elizalde,
{\it Ten Physical Applications of Spectral Zeta Functions}
(Springer-Verlag, Berlin, 1995); A.A. Bytsenko, G. Cognola, L. Vanzo 
and S. Zerbini, Phys. Reports {\bf 266} (1996) 1;
 E. Elizalde, Commun. Math. Phys. {\bf 198} (1998) 83;
V. Moretti, Commun. Math. Phys. {\bf 201} (1999) 327;
V. Moretti, J. Math. Phys. {\bf 40} (1999) 3843;
E. Elizalde, J. Comput. Appl. Math. {\bf 118} (2000) 125.


\bibitem{chao1} J. Main. Phys. Rep. {\bf 316} (1999) 233;
H. Montgomery, {\it Proc. Int. Congress of
Mathematics}, Vancouver (1974), vol. 1, 379;
M. Berry and J. Keating. SIAM Review {\bf
41} No. 2 (1999) 236; A. Connes, Selecta Math. (NS) {\bf 5}
(1999) 29.

\bibitem{sp1} L. de Branges, J. Funct. Anal. {\bf 107} (1992) 122;
J. Alc\'antara-Bode, Int. Eqs. Op. Theory {\bf 17} (1993) 151;
L. de Branges, J. Funct. Anal. {\bf 121} (1994) 117; 
X.-J. Li, Math. Nachr. {\bf 166} (1994) 229;
N. Nikolski, Ann. Inst. Fourier (Grenoble) {\bf 45} (1995) 143;
V.I. Vasyunin, Algebra i Analiz {\bf 7} (1995) 118 [trans. St. Petersburg
Math. J. {\bf 7} (1996) 405].

\bibitem{zfr0} E. Elizalde, J. Phys. {\bf A27} (1994) L299 and 3775.

\bibitem{eejhep1} E. Elizalde, JHEP {\bf 9907} (1999) 015.

\bibitem{nca1} E. Elizalde, L. Vanzo and S.
 Zerbini,  Commun. Math. Phys.  {\bf 194}  (1998) 613;
 E. Elizalde, G. Cognola and S. Zerbini,
Nucl. Phys. {\bf B532} (1998) 407;
J.J. McKenzie-Smith and D.J. Toms, Phys. Rev. {\bf D58} (1998) 105001;
  A.A. Bytsenko and F.L. Williams, J. Math. Phys.
{\bf 39} (1998) 1075;
N. Evans, Phys. Lett. {\bf B457} (1999) 127;
J.S. Dowker,  {\em On the relevance on the multiplicative anomaly}
  hep--th/9803200 (1998); E. Elizalde, A. Filippi, L. Vanzo and S. Zerbini,
 {\em Is the multiplicative
  anomaly dependent on the regularization?}  hep--th/9804071 (1998);
 E. Elizalde, A. Filippi, L. Vanzo and S. Zerbini,
 {\em Is the multiplicative
  anomaly relevant?}  hep--th/9804072 (1998).

\bibitem{wod1} M. Wodzicki. {\it Noncommutative Residue}, Chapter I,
in {\em Lecture Notes in Mathematics}, Yu.I. Manin, editor, Vol. 1289,  320
  (Springer-Verlag, Berlin, 1987);
 C. Kassel, Asterisque {\bf 177} (1989) 199, Sem.~Bourbaki.


\end{thebibliography}
\end{document}